\documentclass{amsart}
\usepackage{amsmath}
\usepackage{amssymb}
\usepackage{amsthm}
\theoremstyle{plain}
\newtheorem{theorem}{Theorem}[section]

\theoremstyle{definition}
\newtheorem{definition}[theorem]{Definition}
\newtheorem{example}[theorem]{Example}

\usepackage{tikz}
\usetikzlibrary{matrix,arrows,positioning,calc}
\usetikzlibrary{shapes.multipart}
\usepackage{tikz-cd}
\pgfdeclarelayer{background layer}
\pgfsetlayers{background layer,main}
\usepackage{booktabs}

\usepackage{color}
\usepackage{colortbl}
\usepackage{hyperref}

\newtheorem{task}{Task}

\begin{document}

\title{On Constructing Finite Automata by
Relational Programming}
\author{Attila Egri-Nagy$^1$}
\address{$^1$Akita International University, Japan}
\email{egri-nagy@aiu.ac.jp}
\author{Chrystopher L. Nehaniv$^2$}
\address{$^2$University of Waterloo, Canada}
\email{cnehaniv@uwaterloo.ca}

\maketitle

\begin{abstract}
    We consider ways to construct a transducer for a given set of input word to output symbol pairs.
This is motivated by the need for representing game playing programs in a low-level mathematical format that can be analyzed by algebraic tools.
This is different from the classical applications of finite state automata, thus the usual optimization techniques are not directly applicable.
Therefore, we use relational programming tools to find minimal transducers realizing a given set of input-output pairs.
\end{abstract}


\section{Introduction}

A fully calculated game tree is a perfect solution for a deterministic
two-player game of complete information.
Working back from the leaves one minimizes the maximum payoff the opponent can
achieve, using a mini-max strategy as demonstrated by von Neumann in 1928 \cite{vonNeumann1928}.
However, the tree may not be the most suitable representation for all purposes.
Here, we are interested in the algebraic study of games \cite{wildbook}.
Therefore, we prefer a finite state automaton representation, as those define canonically associated transformation semigroups.
These, in turn, can be studied by the hierarchical decomposition tools.
The question naturally arises: how can we create finite state automata for playing games?

First we will formalize the automaton construction problem as a machine learning task. Then we will describe relational programming solutions for this task and compare them with classical algorithms.

\subsection{Playing Go -- possible representations}

As an illustration, let's consider how we would implement a playing engine for the game of Go as an automaton.
We would like to have an automaton that can give the next move based on a given board position.
Therefore, the input of the automaton is a sequence of symbols describing a board position.
The board position can be described in at least two ways, as an \emph{image}, or as a \emph{history}.

We can describe the board position as an image, row by row, or by some traversing curve (e.g., a boustrophedon pattern) to have better locality.
We need at least 3 symbols for the local state (white, black, empty intersection).
The input words have the same length, the number of intersections on the board.
However, in Go we have the \emph{ko} rule to avoid infinite loops in the game, so there cannot be an immediate retaking of a captured stone.
Consequently, we would need to use an extra symbol to denote the position forbidden by the ko rule.
Deep learning AIs use this image-like board position, but they address the ko problem differently.
The previous 8 positions are all part of the input layer \cite{AlphaGo2016}.

We can also define the input word as the history, the sequence of moves leading to the given board position from the empty board.
This representation then would use different lengths of input words, depending on the state of the game.

In both representations, it is easy to see that a simple (but enormous) tree structure can realize a perfect player as an automaton.
The question is how compressible is this structure?
That is why we are looking for a minimal automaton that can realize all the desired mappings.

\subsection{Programming Finite State Automata}

Finite state automata (FSA) form a model of computation, in which the input word is read only once, unlike Turing-machines with freely moveable read/write head.
We can consider FSA, and thus transformation semigroups, as a low-level programming language \cite{tsprog2010}.
We can still write programs for this mathematical definition of a computer.
However, FSA are further away from practical programming languages, since the Turing-machine has the tape as a dedicated memory device.
FSA need to implement memory structure in their state sets.
They can have more optimized and efficient memory handling tailored for a given task.
This flexibility comes at a price: FSA are truly horrible programming languages to solve a problem.
Hence, there is a need for algorithms to generate good FSA, instead of manually constructing them.

\section{Formalizing the lossless machine learning problem}

We define a computational task as a list of input-output pairs.
The goal is to construct an automaton that given an input word produces the corresponding output symbol.
Automata that produce output symbols are called \emph{transducers}.
Here we have a special case that produce only a single output symbol for an input word.

This looks like a machine learning problem, so it is important to emphasize the learning here is \emph{not statistical}.
We want to have an error-free reproduction of the outputs for the input words in the training sets.
The generalization is not a goal, only exact reproduction.

\begin{task}[Lossless machine learning: transducer construction]
For an input alphabet $I$ and an output alphabet $O$, we have a list of $(w,r)$ word-symbol pairs, $w\in I^+$, $r\in O$.
We need to construct a special type of transducer automaton $\mathcal{T}$, such that it will produce $r$ when processing $w$ for all such pairs.
\end{task}

\begin{definition}[Single-output transducer]
$\mathcal{T}=(Q, q_o, I, O, \delta, \omega)$,

$Q$ is the set of states, $I$ is the set of input symbols,

$q_o$ is the initial state (thus $Q\neq\varnothing$)

$\delta: Q\times I\to Q$ is the state transition function

$\omega: Q\to O$ is the partial output function
\end{definition}
Transducer $\mathcal{T}$ realizes a partial function $f:I^+\to O$, where we do not care about the values not defined in the input-output pairs.
There is a fixed initial state, in which the transducer starts reading each word.
For an input word $w=w_1w_2\cdots w_k$ of length $k$, $w_i\in I$, the sequence
of states $t_0,\ldots, t_k$ is a \emph{trajectory}, defined recursively by
$t_0=q_o$, and $t_i=\delta(t_{i-1}, w_i)$ for $1\leq i \leq k$.
Then the output for $w$ is defined by $\omega(t_k)$, determined by the last
state in the trajectory.
If $r=f(w)=\omega(t_k)$, then the transducer produces the correct output for the
$(w,r)$ pair in the specification.
There can be states in trajectories with undefined output (but not as last
state), therefore $|Q|\geq |I|$.

\begin{example} The input-output pairs $00\mapsto  0$, $01\mapsto 1$,$10\mapsto
  1$, $11\mapsto 0$ define the textbook problem of checking parity. Here is a
  transducer realizing these pairs.
  \begin{center}
    \vskip-3em
        \includegraphics[height=.2\textheight]{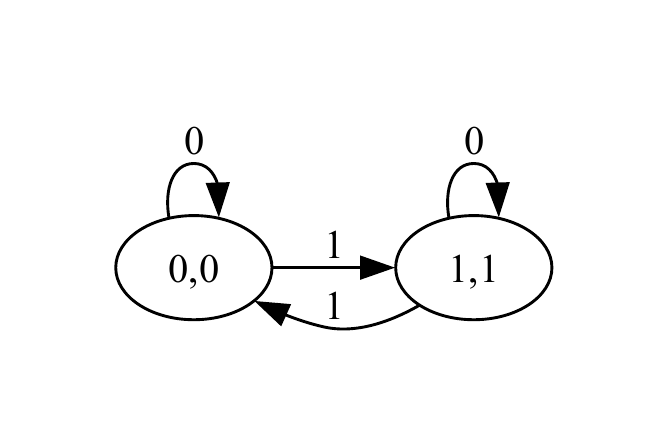}
    \end{center}
    $Q=I=O=\{0,1\}$, $q_0=0$, $\omega$ is the identity function. The arrow labels are the input symbols. Inside a node representing a state, we have the state itself followed by the corresponding output value.  
\end{example}

\subsection{Construction Meta-Algorithms}

There are several factors to consider when planning to solve this task.
First, we do not know in general how many states are needed to realize a given set of input-output pairs.
The minimum number of states is often the ultimate question in a problem domain, like for the game playing transducers.
What is the complexity of a game?
It is the smallest number of states needed for a transducer to be able to produce an optimal move for all board positions.\footnote{Alternatively, one can define it as the Krohn-Rhodes complexity of the minimal automaton. (See \cite{wildbook}.)}
In practice, we can simply start with a single state $|Q|=1$, and add a state when it becomes clear that there is an insufficient number of states.
 Another option would be to build an easy to construct transducer with possibly many redundant states, and apply an algorithm to prune the state set.

Another issue to consider, is the order of the input-output pairs in the construction process, and thus the incremental nature of the algorithm.
In general, we cannot expect a gradual structure evolution.
If we construct a transducer for $n-1$ input-output pairs, adding the $n$th pair may require a completely different transducer.


\section{Relational programming solution}

Finding a correct transducer (not to mention a minimal one) for a set of input-output pairs is a daunting task for a human programmer.
Therefore, it is a natural idea to automate it.

We use logic programming, or more generally relational programming.
The basic idea can be illustrated with an elementary mathematical example.
In functional thinking, we consider exponential and logarithmic functions as functions that produce outputs from input values, e.g., $f(x)=2^x$ with input 3 we get output 8, since $f(3)=2^3=8$.
We can also consider its inverse function, where the input 8 gives output 3, since $\log_2 8=3$.
In relational thinking there is no input, there is no output. Just three numbers in a relationship: $r(2,3,8)$.
If any of the numbers missing, we can recover that value: $r(2,?,8)$ will give 3, and $r(?,3,8)$ will give 2.
  
We can have the same distinction for the problem of transducer construction.
Functionally speaking, a transducer produces an output symbol from a sequence of input symbols.
Relationally speaking, we have a relationship between input, output and the transducer.
Therefore, if the input-output pairs are given, we can infer the transducer.
The inference is done by a search algorithm for finding variable combinations satisfying the constraints of the relation.

\subsection{Implementation}
We used the \texttt{miniKanren} \cite{MiniKanren} family of programming
languages for relational programming.
It has a minimalistic approach famously described by only two pages of source
code in \cite{ReasonedSchemer}.
This simplicity also means that it is easy to implement, thus there are  several implementations in many programming languages.
In particular, we use the \texttt{core.logic} library written in \texttt{Clojure} \cite{Clojure2020}.
The code for the transducer construction algorithms below can be found in our \texttt{kigen} package \cite{kigen}.
In this project, we treated the logic engine as a black box, i.e., we did not
consider any internal parameters of the search.
We simply defined the logic variables and their constraints.

\subsection{Logic Variable Considerations \&  Setup}

When using a logic search engine, the main question is `What are the logic variables?'. 
The constraints are the $(w_i,r_i)$ pairs, the unknown part is the transducer itself.
We define the state set to be non-negative integers $Q=\{0,1,2,\ldots,n-1\}$.
This is fixed because the states are used as indices internally in the state transition table.
The initial state is set to $q_0=0$.
If we leave the output map $\omega$ flexible, then this fixed choice does not affect the resulting transducer.
The input and output symbols can be any distinct symbols without restrictions (anything the underlying language permits). 
We also need the flexibility for the output function.
We cannot fix the outputs to be the first $n$ states as that would impose an additional constraint with the fixed initial state.

The development of the software implementation proceeded from the hard-coded representation (states, input and output symbols represented by non-negative integers) to a more flexible, symbolic hash-map based data structure, as it became clear that fixed choices interfere with satisfiability with a given number of states.

The logic variables are associated with the entries of the state
transition table  $Q\times I$ under $\delta$, and with the output values of the
states under $\omega$.
Thus, the number of logic variables is
$|Q|(|I|+1)$.
It also became inevitable during the development to allow for partial $\delta$ and $\omega$.
Whenever a state transition is not defined, the value \texttt{nil} is used.
This indicates that the mapping is not needed.
For $\omega$, there need to be at least $|O|$ entries defined.
However, adding another possible value for the logic variables increases their number.
Thus, it is more efficient to let the logic engine come up with a solution where the functions are totally defined and then simulate the transducer on all input words and record which mappings are not used.
The number of defined maps in the state transition table is also a measure for resource usage.
If there are many entries undefined, it is likely that task can be solved by a smaller number of states.

Another variation is also considered.
Each word $w$ defines a trajectory $q_0,t_1,\ldots,t_k$, a sequence of states starting with the initial state.
We can define logic variables for each $t_i$ state in all trajectories.
The constraints are the properties of functions, i.e., the state transitions
labeled by the input symbols and extracted from the trajectories should define a
valid $\delta$ function.
In other words, if an input symbol takes state $q_i$ to $q_j$ in one place, then
all other occurrences of $q_i$ in all trajectories should go to $q_j$ under the
same input symbol.
This method gives partial solutions directly (no need for simulations).
However, the number of logic variables grows with the total length of the input words, and finding solutions quickly becomes computationally unfeasible.

\subsection{Examples}

Here we describe some examples of the constructed transducers.

\begin{figure}
    \includegraphics[width=\textwidth]{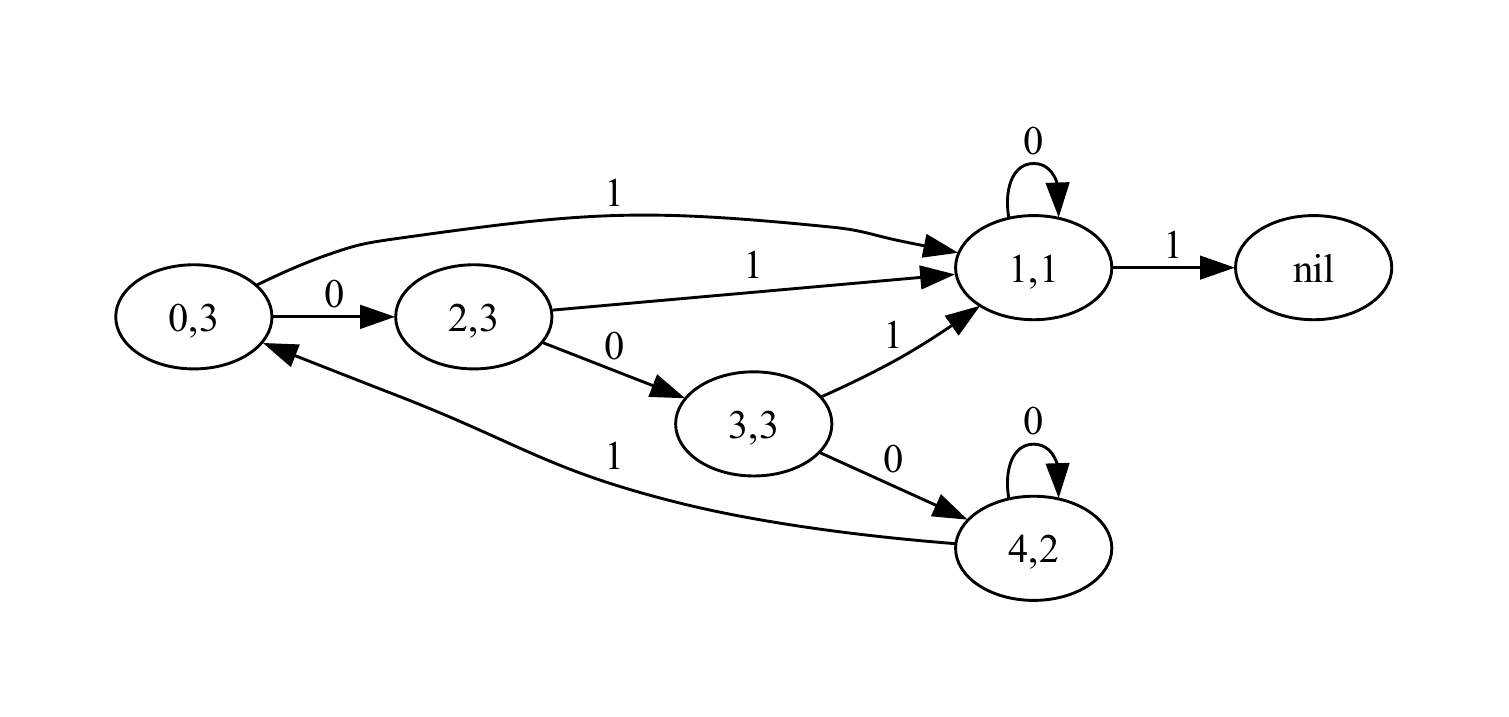}
    \caption{\textbf{Signal Locator 9-3.} The transducer for locating 1 among the zeroes in a bit-string of length 9. The node named nil is a sink state, representing partial transformations. We can see what allows the state transition to remain undefined. State 1 represents the case when the signal is in the first third, thus no more ones can be read. To detect the signal in the second third, we need to reach state 4, then return to the initial state and back to 4 `in time'. To locate the signal in the last third, we do the same, just not getting back to 4. This transducer would be easy to design manually, but it is also interesting enough to study the algorithm of this minimal solution.}
    \label{fig:sl-9-3}
\end{figure}

\begin{example}[Signal Locator 9-3]
$I=\{0,1\}$, the input words are bit-strings of length 9, all zeroes except one symbol. The task is to find in which third of the input number 1 (the signal) occurs.
For instance, 000010000 $\mapsto$ 2.
This can be done with 5 states (see Fig.~\ref{fig:sl-9-3}), but not with 4
states. The size of the search space is $5^{10}\cdot 3^5$ for the 15 logic
variables.
The task can be generalized to $n$-$k$, locating the signal in $n$-bit sequences
partitioned into $k$ parts.
\end{example}

\begin{figure}
\begin{center}
    \includegraphics[height=.21\textheight]{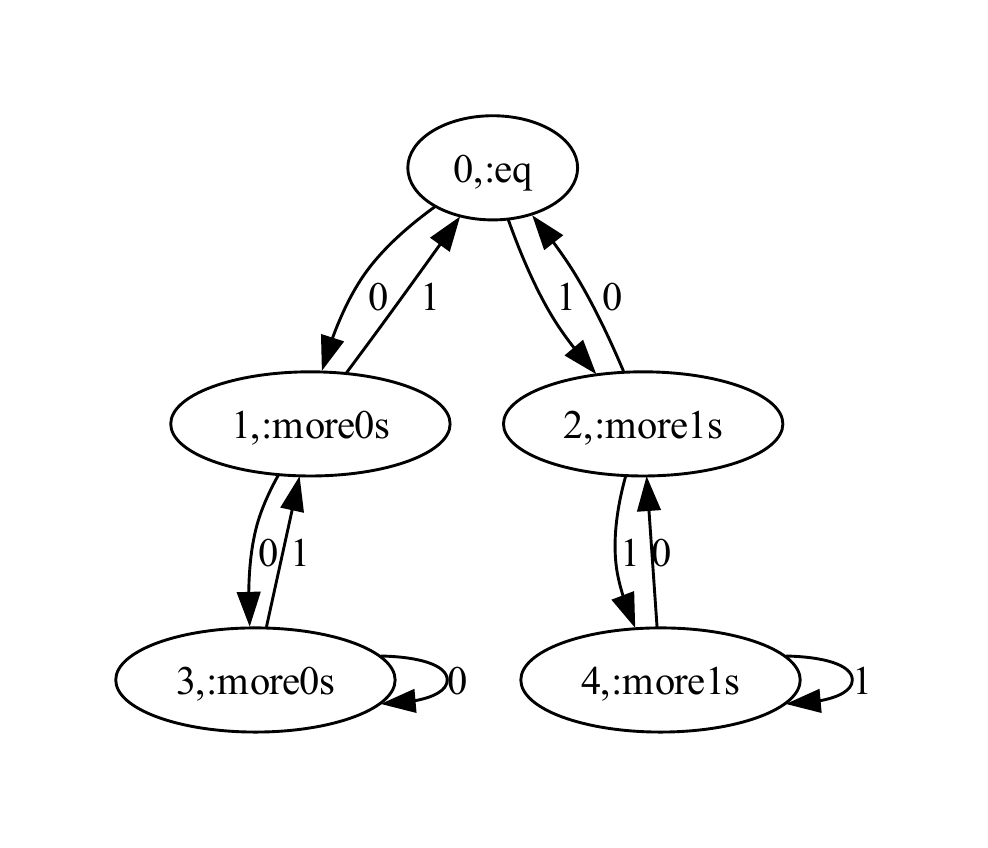}    
    \includegraphics[height=.38\textheight]{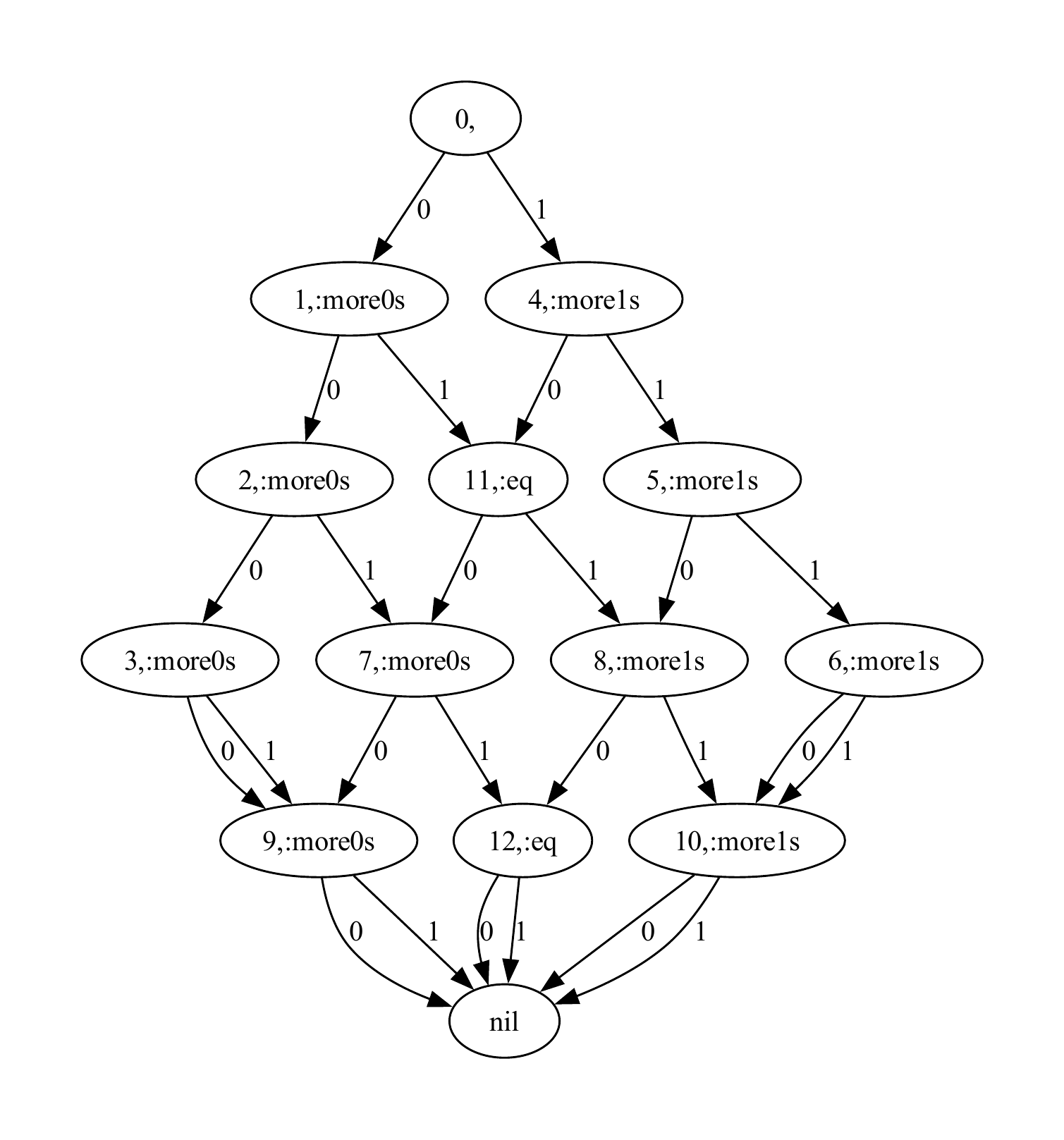}    
\end{center}
    \caption{\textbf{Zeroes or ones.} A solution for the `Zeroes or ones' problem for bitstrings of length 4. Left: minimal one by the logic engine, right: minimized trie construction. The difference is clear: the minimized trie has only forward edges. }
    \label{fig:zo4}
\end{figure}

\begin{example}[Zeroes or ones]
The input words are bitstrings of a given length and the transducer has to
decide whether the number of zeroes is equal to, less or more than the number of
ones.
The task is similar to the majority function in Boolean logic.
The solution is essentially a bi-directional counter (see Fig.~\ref{fig:zo4}).
Search space size is $5^{10}\cdot 3^5$ for the 15 logic variables.
\end{example}

\begin{figure}
    \includegraphics[height=.32\textheight]{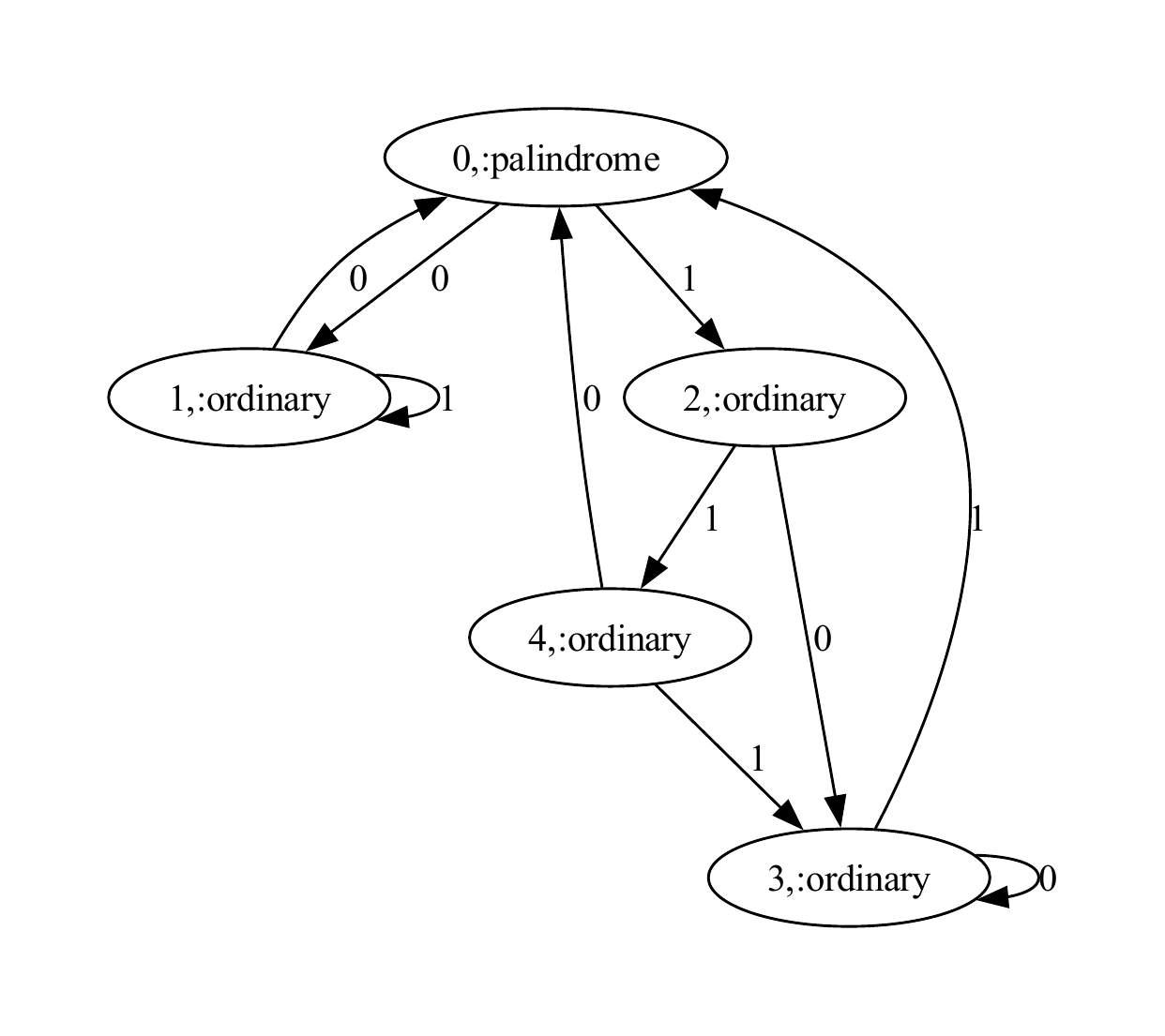}
    \includegraphics[height=.45\textheight]{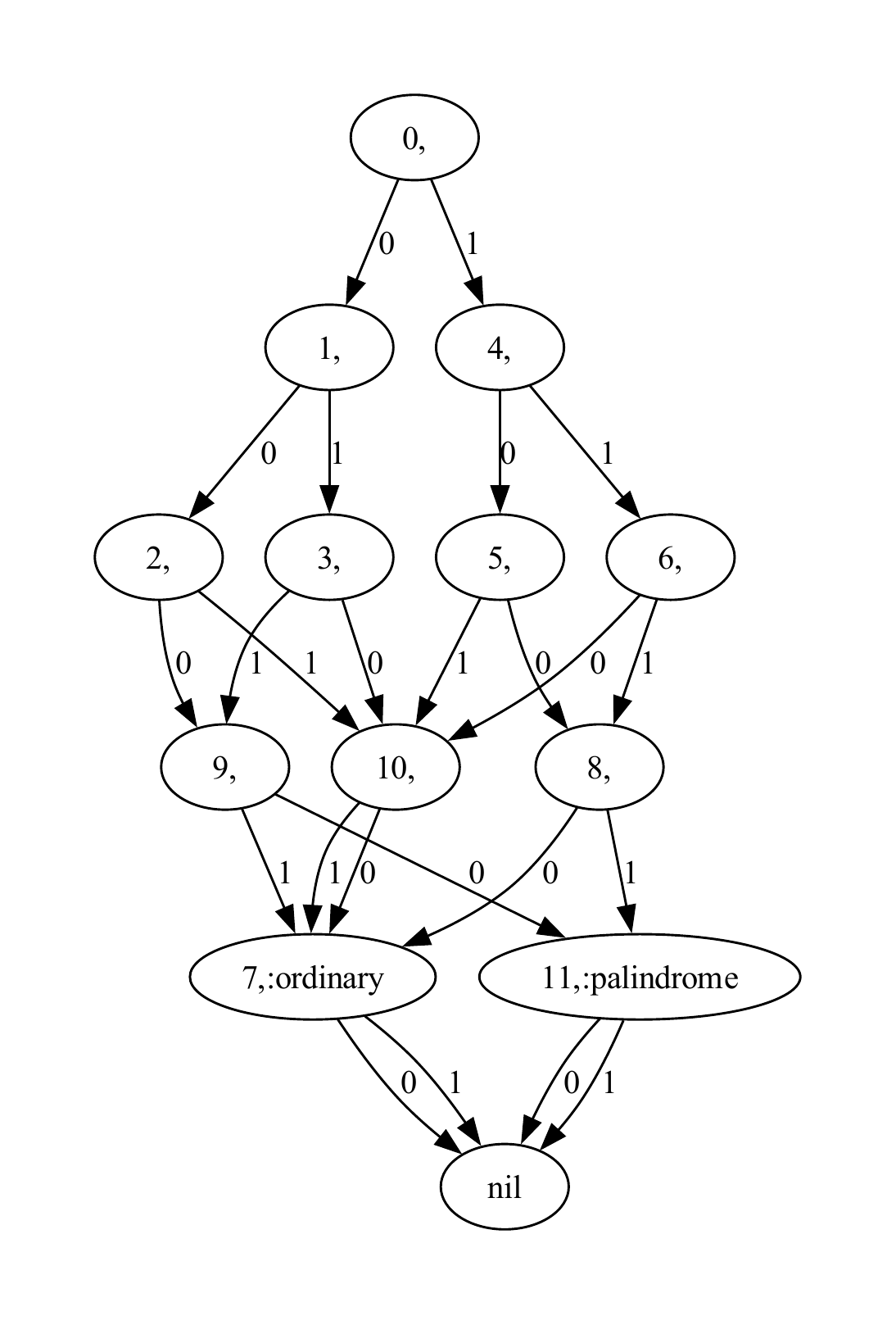}
    \caption{\textbf{Palindromes.} Transducers for detecting palindromic bitstrings of length 4; minimal and classic trie-based version.}
    \label{fig:palindrome}
\end{figure}

\begin{example}[Palindromes]
Recognizing palindromes (words that read the same backwards) is the prime example of a problem that cannot be solved by FSA. The language of palindromes is context-free, but not regular.
However, for a given fixed length, we can construct a transducer for deciding
this property. For length 4 binary strings the search space size is $5^{10}\cdot
2^5$, and it is easy to find a minimal solution (Fig.~\ref{fig:palindrome}).
However, the logic engine solution does not scale well. Length 5 is not an
immediately finishing calculation.
\end{example}

\begin{example}[Word classification]
The task is to sort natural language words into three groups: English (eruption, erudite,
oriental, topology, serendipity), Japanese (eki, origami, arigato), and
Hungarian (asztal, mester).
The search space size is $3^{54}$.
There is a solution with three states.
\end{example}


\section{Comparison with classical minimization algorithms}

A remarkable result of automata theory is that every FSA has a unique minimal (regarding the number of states) equivalent automaton.
This seminal result suggests that we have a simple solution for our transducer construction problem.
We could simply construct any transducer satisfying the input-output constraints and minimize that automaton.
However, satisfying the input-output constraints only is a more general problem on two accounts.
First, recognizing words is a special case of classifying them.
Second, the crucial point here, we have no constraints on subwords of input words, and on words others than those specified in the input-output pairs.
Therefore, we do not get a minimal transducer by the classical methods.

For applying the classical minimization algorithms, we need to establish that there is no fundamental difference between an automaton used as a recognizer and as a single-output transducer.
The latter case can be seen as a more nuanced recognizer.
The output symbols can be seen as different classes of acceptance.
Or, the other way around, a recognizer is a single output transducer with only two output symbols (accept and reject).

To construct a working transducer we created a search \emph{trie}, which is a tree encoding the words in the branches with the shared prefixes represented by a single path.
Then we used the standard minimization algorithm based on the Myhill-Nerode Theorem \cite{HopcroftUllman79}.
The algorithm constructs an increasingly finer partition of the state set.
Two states are equivalent if they yield the same output symbol for the same input word.
Initially, we have the equivalence classes for the input symbols at the leaf nodes of the trie, and a class for all the undefined output for internal nodes.
When we split an equivalence class, the states in other classes need to be checked for whether this new distinction can tell them apart.
When there is no more splitting, the equivalence classes define the states of the minimal transducer.

This classical trie-minimization method is lot faster than the search algorithm in the logic engine, but in terms of number of states used it is far from the minimal values.
Here is a comparison for the examples above.
\begin{center}
\begin{tabular}[pos]{lccc}
    \toprule
Task  & Minimal & Trie  & Minimized\\
\midrule
Signal Locator 9-3 & 5 & 45 & 24 \\ 
Signal Locator 8-4 & 6 & 36 & 23 \\
Zeroes and ones 4 & 5 & 31 & 13\\
Palindrome 4 & 5 & 31 & 12 \\
Word Classification & 3 & 68 & 57\\
    \bottomrule
\end{tabular}
\end{center}
The main reason for the difference is that the trie method does not have the
flexibility for assigning the output values to states.
The word classification demonstrates that the prefixes are shared by the trie
construction, and by optimization some word endings can be identified
too.
However, it is still about tracing the individual words letter by letter, hence
the large number of states.
Since we do not care about classifying subwords, the relational programming
search method can find a  minimal possible transducer.

\section{Conclusion and Future Work}

We applied relational programming to the problem of constructing minimal transducers for a given set of pairs consisting of  input words and their corresponding output values.
We compared the method with the classical method of minimizing a transducer constructed by a trie.
In summary, the relational method produces a truly minimal transducer, but it is computationally intensive and does not scale well.
In contrast, the traditional method does not yield a minimal transducer.
This does not contradict the fundamental theorem about the unique minimal automaton for any deterministic finite state automaton, since we have flexibility in defining the output function, and we do not care about values associated with input words not in the training set.

There are at least two different ways for trying to  further advance solution of  this non-statistical machine learning problem:
\begin{enumerate}
    \item Developing an incremental construction algorithm using relational programming;
    \item Adopting more advanced classical approaches (e.g., \cite{AhoCorasick1975}) to the flexible transducer construction.
\end{enumerate}

\section*{Acknowledgements}
This project was funded by the Kakenhi grant
22K00015 by the Japan Society for the Promotion of Science (JSPS), titled `On progressing human understanding in the shadow of superhuman
deep learning artificial intelligence entities' (Grant-in-Aid for Scientific
Research type C, \url{https://kaken.nii.ac.jp/grant/KAKENHI-PROJECT-22K00015/}).
This research was also supported in part by the Natural Sciences and Engineering Research Council of Canada (NSERC), funding reference number RGPIN-2019-04669. 
Cette recherche a \'et\'e financ\'ee en partie par le Conseil de recherches en sciences naturelles et en g\'enie du Canada (CRSNG), num\'ero de r\'ef\'erence RGPIN-2019-04669.

\bibliographystyle{plain}
\bibliography{../coords}

\end{document}